\documentclass[lettersize,journal]{IEEEtran}
\usepackage{amsmath,amsfonts}
\usepackage{algorithmic}
\usepackage{algorithm}
\usepackage{array}
\usepackage{booktabs}
\usepackage[caption=false,font=footnotesize,labelfont=rm,textfont=rm]{subfig}
\usepackage{textcomp}
\usepackage{stfloats}
\usepackage{url}
\usepackage{verbatim}
\usepackage{graphicx}
\usepackage{float} 
\usepackage{cite}
\usepackage{amsthm}
\usepackage{amssymb}
\usepackage{xcolor}
\usepackage{dsfont}
\usepackage{soul}
\usepackage{makecell}
\usepackage{bbding}
\newlength \figwidth
\definecolor{bittersweet}{rgb}{1.0, 0.44, 0.37}
\definecolor{glaucous}{rgb}{0.38, 0.51, 0.71}
\definecolor{gainsboro}{rgb}{0.86, 0.86, 0.86}
\definecolor{babyblueeyes}{rgb}{0.63, 0.79, 0.95}
\definecolor{silver}{rgb}{0.75, 0.75, 0.75}
\definecolor{neoncarrot}{rgb}{1.0, 0.64, 0.26}
\definecolor{Gray}{gray}{0.6}
\definecolor{LightCyan}{rgb}{0.88,1,1}
\definecolor{BackgroundLightBlue}{rgb}{0.97,0.97,1}
\definecolor{BackgroundGray}{gray}{0.98}

\def\nb0{{\mathbf{0}}}
\def\nb1{{\mathbf{1}}}

\begin{document}

\bstctlcite{IEEEexample:BSTcontrol}


\title{Non-Terrestrial Networking for 6G: Evolution, Opportunities, and Future Directions}

\author{
Feng Wang,
Shengyu Zhang,
Huiting Yang,
and Tony Q. S. Quek

\thanks{Feng Wang, Shengyu Zhang, Huiting Yang, and Tony Q. S. Quek are with the Information System Technology and Design Pillar, Singapore University of Technology and Design, Singapore 487372 (e-mail: feng2\_wang@sutd.edu.sg; shengyu\_zhang@sutd.edu.sg; huiting\_yang@sutd.edu.sg; tonyquek@sutd.edu.sg). (Corresponding author: Tony Q. S. Quek.)}
}



\maketitle

\begin{abstract}
From 5G onwards, Non-Terrestrial Networks (NTNs) have emerged as a key component of future network architectures. Leveraging Low Earth Orbit (LEO) satellite constellations, NTNs are capable of building a space Internet and present a paradigm shift in delivering mobile services to even the most remote regions on Earth. However, the extensive coverage and rapid movement of LEO satellites pose unique challenges for NTN networking, including user equipment (UE) access and inter-satellite delivery, which directly impact the quality of service (QoS) and data transmission continuity. This paper offers an in-depth review of advanced NTN management technologies in the context of 6G evolution, focusing on radio resource management, mobility management, and dynamic network slicing. Building on this foundation and considering the latest trends in NTN development, we then present some innovative perspectives to emerging challenges in satellite beamforming, handover mechanisms, and inter-satellite transmissions. Lastly, we identify open research issues and propose future directions aimed at advancing satellite Internet deployment and enhancing NTN performance.

\end{abstract}

\begin{IEEEkeywords}
Non-Terrestrial Networks (NTN), 6G, networking, satellite beamforming, mobility management, inter-satellite transmission.
\end{IEEEkeywords}

\section{Introduction}

\subsection{Background}

From 5G and beyond, Non-Terrestrial Networks (NTN) have become a crucial component of future network architectures, playing a pivotal role in achieving the vision of three-dimensional global seamless coverage \cite{bibitem1,bibitem2}. NTN, which include Geostationary Orbit (GEO) satellites, Medium Earth Orbit (MEO) satellites, Low Earth Orbit (LEO) satellites, and high-altitude platforms (HAP), are capable of providing network connectivity to any corner of our Earth. This capability is especially vital in remote areas where terrestrial network (TN) coverage is insufficient or economically unfeasible, such as mountainous regions or oceans~\cite{bibitem3}. Moreover, in scenarios where TN infrastructures are compromised by natural disasters or emergencies, NTN serves as an essential connectivity alternative~\cite{bibitem4}. With the advancement of onboard data processing capabilities and the development of large constellations, satellite communication is becoming a prominent technology in the evolution of NTN, attracting significant attention from both industry and academia~\cite{bibitem5,bibitem6}.

The development of LEO constellations is spearheading the evolution of NTN construction due to their proximity, ranging from 500 km to 1200 km above the Earth, which enables low-latency network connectivity and enhanced radio link budgets compared to higher orbit satellites~\cite{bibitem7,bibitem8}. The evolution of LEO communication technologies can be characterized in three distinct phases, each also reflecting advancements in user terminal (UE) types:
\begin{itemize}
    \item The first phase involved dedicated satellite systems where UE operated independently with both TN and NTN, typically like a satellite \& cellular hybrid dual-mode phone that was often large in size.
    \item The current second phase involves installing a cellphone tower in space. This allows UE to operate without hardware or firmware modifications, enabling NTN to provide direct-to-device services and support Sub-6 GHz cellular data services.
    \item The anticipated third phase, named 5G NTN enhancement, will involve new designs on both satellite and UE to further boost NTN service capabilities and reliability, aiming for a unified TN-NTN constructions.
\end{itemize}
Historically, these phases also mirror the evolving relationship between TN and NTN: starting from independent development with interworking before 5G, moving to TN innovation with minimal changes to support NTN in 5G and 5G-Advanced, and evolving towards a unified TN-NTN design and optimization in 6G and beyond. Regarding NTN development in standardization activities, the 3GPP work in Rel-15 and Rel-16 established key NTN use cases and system architectures while providing satellite-based NTN channel models and system-level simulation assumptions. In Rel-17, 3GPP identified NTN as part of the 5G NR ecosystem, detailing NTN use cases for sub-6 GHz frequencies, including massive access and narrowband IoT applications. Rel-18 explored enhancements for 5G NR NTN operations for handheld terminals on 10 GHz and above, focusing on spectrum management and co-existence to facilitate efficient integration with TN~\cite{bibitem9,bibitem10,bibitem11,bibitem12}. As ongoing discussions in Rrel-19, NTN is considered a long-term study area, pointing towards its evolving and crucial role in future telecommunications frameworks. Following the NTN development trajectory, building a reliable space Internet through strategic networking technology to connect every corner of the world will soon be realized.

\subsection{Challenges and Motivation for NTN Networking}

For LEO-based NTN, the extensive coverage and dynamic features offer opportunities for ubiquitous access. However, they also present unique challenges for efficient networking in NTN, particularly in the management of radio resources, system mobility, and onboard traffic scheduling, detailed as follows.

\subsubsection{Radio resource management in NTN}
The dense access in NTN often leads to competition among numerous UEs for limited bandwidth, leading to inevitable signal interference issues~\cite{bibitem13,bibitem14,bibitem15,bibitem16,bibitem17}. Effective interference management is crucial to maintaining the reliability of NTN accessing~\cite{bibitem18,bibitem19,bibitem20}. The challenge primarily involves accurately estimating Channel State Information (CSI) and developing efficient beamforming strategies~\cite{bibitem21,bibitem22,bibitem23}. CSI provides crucial details about the current state of communication channels, typically assessed based on known signals received at the UE and subsequently relayed to the transmitter for use in future transmissions. Accurate CSI is essential as it allows for informed resource allocation decisions and the formulation of strategies to mitigate signal interference. However, accurately acquiring CSI is complex in NTNs due to the rapid movement of LEO satellites~\cite{bibitem24,bibitem25,bibitem26}. Traditional CSI estimation techniques relying heavily on receiver feedback face challenges in NTN like latency, resulting in potentially outdated CSI~\cite{bibitem27}. Additionally, this challenge is exacerbated by the delay in processing CSI for beamforming design, which must meet Quality of Service (QoS) requirements while managing interference. Conventional methods such as Successive Convex Approximation (SCA)~\cite{bibitem28,bibitem29} and Weighted Minimum Mean Square Error (WMMSE)~\cite{bibitem30} offer solutions but are computationally demanding and sensitive to initialization choices, leading to further delays. These intertwined challenges underline the complexity of radio resource management in NTN and emphasize the need for more adaptive and efficient CSI estimation and beamforming techniques.

\begin{figure}[!t]
\centering
\includegraphics[width=0.95\figwidth]{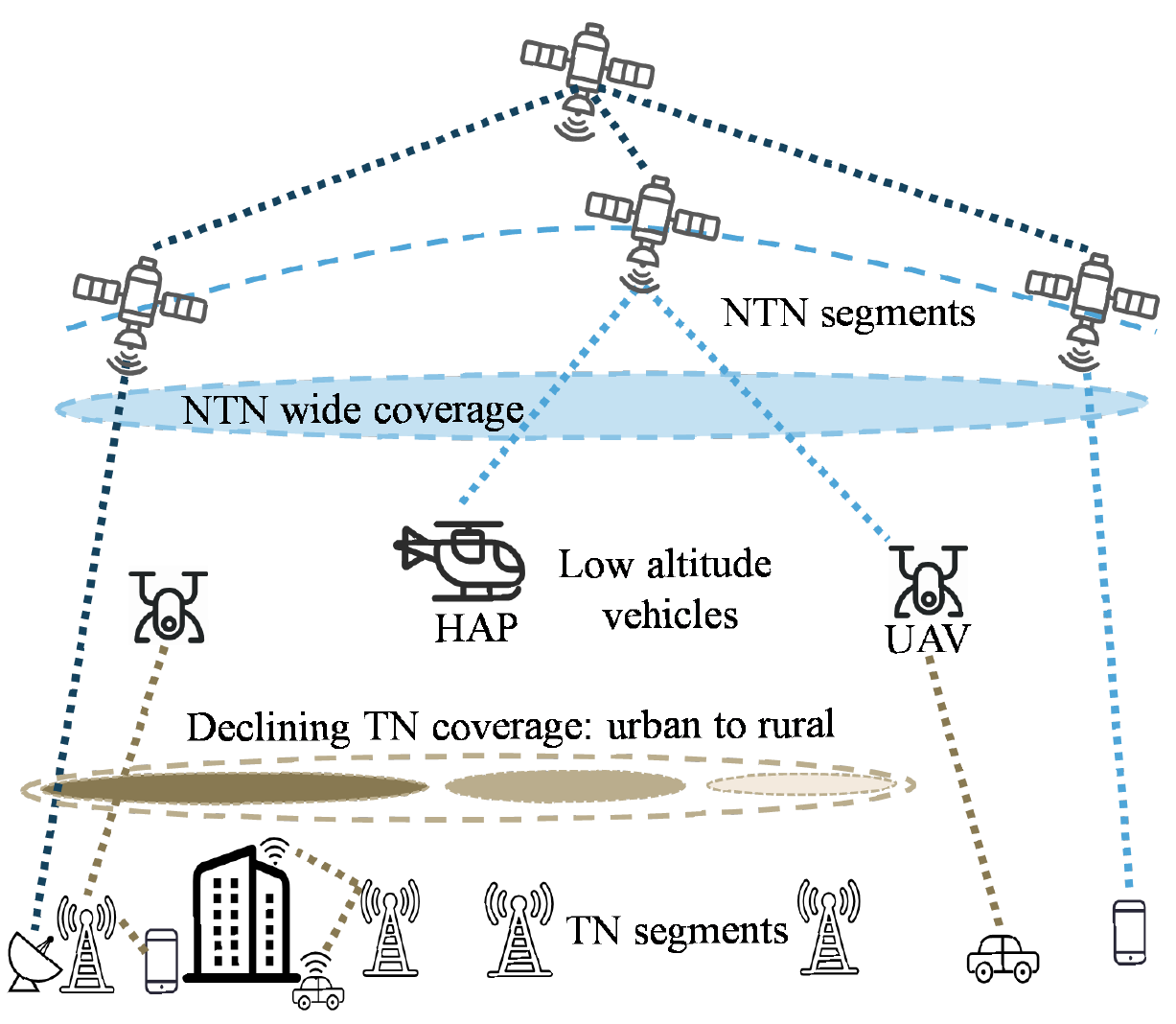}
\caption{NTN diverse networking scenarios and comparison with TN.}
\label{fig1}
\end{figure}

\subsubsection{Mobility management in NTN}
For mobility management in LEO-based NTN, traditional TN handover solutions are insufficient due to the distinct dynamics of LEO satellites~\cite{bibitem7}. Unlike ground BSs, LEO satellites move at high speeds, leading to frequent handovers and fluctuating signal strengths, influenced by the elevation angle and additional atmospheric losses. Compared to TN, NTN also experiences more severe Doppler shifts due to the rapid movement of satellites. Coverage by a single LEO satellite might last only seconds to minute, necessitating timely handover processing, including signaling and synchronization. Moreover, as TN and NTN integration progresses, mobile UEs like cars and UAVs may travel from a TN dense area to a sparse area where both TN and NTN cell coverage exist, as shown in Fig. 1. Under unified spectrum management, UEs are allowed to handover from TN to NTN for a better data reception. It requires addressing both intra-NTN mobility and mobility management between TN and NTN to ensure service continuity. Therefore, NTN mobility management poses a crucial challenge in its ongoing evolution.

\subsubsection{Traffic scheduling in NTN}
Current NTN systems, with multiple constellations, often face significant limitations in flexibility and adaptability due to constellation being custom-built for specific tasks, such as Earth observation or broadcasting. This leads to underutilized onboard resources, as these satellites cannot easily be re-connected for other tasks during idle periods~\cite{bibitem31,bibitem32,bibitem33,bibitem34,bibitem35,bibitem36,bibitem37,bibitem38,bibitem39}. To overcome these limitations, virtualization technologies like Network Function Virtualization (NFV) and Software-Defined Networking (SDN) have been introduced, creating Software-Defined NTNs (SD-NTNs) for centralized inter-satellite networking and efficient resource scheduling~\cite{bibitem40,bibitem41,bibitem42,bibitem43}. A critical feature enabled by SD-NTNs is network slicing, which allows for the creation of multiple isolated virtual networks, known as network slices, over a shared physical infrastructure~\cite{bibitem44,bibitem45,bibitem46}. Each network slice is tailored to meet the specific requirements of different services or applications, such as enhanced mobile broadband (eMBB), ultra-reliable low-latency communication (URLLC), and massive machine-type communication (mMTC)~\cite{bibitem45,bibitem46,bibitem47,bibitem48}, significantly enhancing operational efficiency, responsiveness, and resource utilization for NTN. However, network slicing in SD-NTNs faces challenges due to the dynamical motion, resource constraints, and diverse service demands~\cite{bibitem49,bibitem50,bibitem51,bibitem52}. Efficient resource allocation and the resolution of service conflicts are critical for optimizing NTN networking, ensuring resource efficiency, and supporting diverse applications.

In the following, we first summarize recent NTN advances with a focus on radio resource management, system mobility, and onboard traffic scheduling. We aim to provide a comprehensive overview of the evolution of NTN networking technology in satellite access, handover processes, and onboard transmissions, while highlighting some innovative perspectives on these elements. Finally, we identify some open problems and future research directions that could foster significant breakthroughs in NTN networking.

\section{Related Advancement}

\subsection{CSI Estimation}
To cope with the outdated CSI, many researchers have focused on the long-term estimation or prediction of the CSI. The CSI prediction can be categorized into two classes: stochastic modelling and Artificial Intelligence (AI)-based methods. The conventional stochastic methods relied on the statistical modelling of wireless channels to predict the CSI. The most widely used method is the AutoRegressive (AR) model, which uses a weighted linear combination of the historical CSI data to forecast the future CSI and approximates the channel as an AR process~\cite{bibitem53,bibitem54}. However, the AR model can only enable one-step prediction and incurs huge computational complexity. To tackle this problem, neural networks, with their inherent ability to capture temporal dependencies, are emerging as a powerful new paradigm for accurate and adaptive channel prediction. In~\cite{bibitem55}, based on MultiLayer Perceptron (MLP), Turan et al. have proposed a novel framework for channel modelling in V2X. In~\cite{bibitem56}, a Convolutional Neural Network (CNN)-based predictor is proposed to improve the prediction quality for high mobility scenarios. Ding et al. first consider the Recurrent Neural Network (RNN) for channel prediction in a fading environment~\cite{bibitem57}. Taking advantage of the memory mechanism, the Long Short-Term Memory (LSTM) achieves better performance in predicting sequential data. In~\cite{bibitem58}, Peng et al. proposed an LSTM-based framework for predicting the channel in massive Multiple-Input and Multiple-Output (MIMO) communications under imperfect CSI. Recently, an attention-based deep learning approach, known as the transformer, has been designed for processing sequential data. It has been proved that the transformer-based structure can significantly enhance the accuracy of channel prediction~\cite{bibitem59}. 

\subsection{Beamforming Design}
Precoding plays a pivotal role in enhancing the quality of transmitted signals by mitigating interference and improving the spectrum efficiency. One of the conventional precoding techniques is Zero-Forcing (ZF) precoding, which aims to cancel out interference at the receivers~\cite{bibitem60}. However, the performance of ZF precoding is known to be sensitive to the quality of CSI. To address this limitation, iterative algorithms such as Successive Convex Approximation (SCA)~\cite{bibitem28,bibitem29} and Weighted Minimum Mean Square Error (WMMSE)~\cite{bibitem30} have been proposed for optimizing precoders under imperfect CSI conditions. However, due to the substantial number of users and antennas in NTNs, the computational complexity of these methods can be prohibitively high. Then, some of the incentive approaches are proposed to solve the precoding design efficiently, and motivate the UTs to participate the interference management~\cite{bibitem13}. Inspired by the superiority of Deep Learning (DL) for solving non-convex problems, researchers have introduced DL-based approaches for efficient precoder design. For example, Liu et al. introduced a DL-based approach for precoder design in Vehicle-to-Infrastructure (V2I) networks~\cite{bibitem23}, which was later extended to address Unmanned Aerial Vehicle (UAV) communications~\cite{bibitem61}. Additionally, a Long Short-Term Memory (LSTM)-based precoding framework has been proposed~\cite{bibitem62}, with subsequent applications in vehicular networks~\cite{bibitem63}. These DL-based approaches offer promising solutions for optimizing precoder design in communication systems, potentially reducing computational complexity while improving performance. 

\subsection{NTN Mobility Management}

\subsubsection{From TN to NTN}
Regarding NTN mobility management, LEO satellite constellations create overlapping coverage areas, presenting multiple handover candidates for each UE, which calls for more efficient HO strategies~\cite{bibitem64}. Traditional TN handover strategies based on radio quality have been adapted for NTN~\cite{bibitem65,bibitem66,bibitem67}, resulting in high HO rates due to minimal signal variation. Studies have investigated how NTN capacity and throughput change with satellite-UE distance, proposing threshold triggers to enhance HO effectiveness~\cite{bibitem68}. Additionally, NTN-specific handover criteria have been developed based on satellite and UE positions, utilizing predictable satellite orbital movement to reduce handover overhead~\cite{bibitem69,bibitem70}.

\subsubsection{Mobility management with network optimization}
Optimizing target satellite selections to boost NTN service performance and efficiency is crucial. Researchers have incorporated satellite HO strategies into multi-metric network optimizations to refine radio resource allocation~\cite{bibitem71,bibitem72,bibitem73} or facilitate traffic offloading~\cite{bibitem74,bibitem75}. For example, Ji et al.~\cite{bibitem8} developed a dynamic NTN management architecture for global information acquisition, supporting satellite HO schemes aimed at minimizing signaling overhead and transmission delays. NTN HO decisions are also treated as multi-objective optimization problems to reduce HO failure rates~\cite{bibitem76}. Additionally, network flow models have been utilized to optimize satellite HOs, ensuring efficient resource use and reducing conflicting selections~\cite{bibitem74,bibitem77}. Furthermore, tailored HO strategies have been developed to meet diverse UE requirements, employing approaches like game theory to derive HO decisions maximizing UE satisfaction~\cite{bibitem78}. 

\subsubsection{NTN service continuity advancements}
Focus has also been placed on extending the time of stay of satellites through optimized NTN HOs for a stable data service. Al-Hourani developed an analytical model to estimate session durations between consecutive HOs~\cite{bibitem79}, further influencing the strategic deployment of LEO constellations~\cite{bibitem80}. Predictive HO decisions have been introduced to extend service durations and minimize disruptions~\cite{bibitem81,bibitem82,bibitem83}. Machine learning has proven instrumental in heterogeneous networking for optimizing HO decisions within NTN, enhancing system responsiveness and reliability~\cite{bibitem84,bibitem85,bibitem86}. Moreover, research is exploring HO strategies between TN and NTN to support seamless mobility for moving UEs like cars, high-speed trains, and UAVs, thus enhancing sustainable and high-quality data reception in integrated systems~\cite{bibitem87,bibitem88}.

Overall, while existing works focus on specific network performance optimizations within NTN mobility management, they often divide the whole service duration to discrete snapshots to decide each HO target sequentially. This generally overlooks the interdependencies between consecutive HOs and potential better future selections. Each HO decision is interconnected, potentially leading to various satellite handover sequences and influencing the overall system service continuity. Additionally, effective TN-NTN handover strategies must consider the distinct coverage and signal variations between TN and NTN.

\subsection{SD-NTN and Traffic Scheduling}
Recently, the SD-NTN architecture has been intensively investigated~\cite{bibitem34,bibitem37,bibitem43,bibitem89,bibitem90}. In ~\cite{bibitem34}, a cross-domain SDN architecture for the multi-layered space-terrestrial integrated network was proposed to enhance flexibility and efficiency in managing diverse network domains, with improvements in configuration updates and decision-making processes. In~\cite{bibitem7}, a software-defined broadband satellite network architecture was investigated to enable flexible and scalable resource management, and an optimal allocation strategy was proposed to improve resource sharing and cooperation among various network resources. In~\cite{bibitem43}, a software-defined space-terrestrial integrated network architecture was proposed to enhance network management, flexibility, and service quality. In~\cite{bibitem89}, a software-defined space-air-ground integrated vehicular network architecture was proposed to support diverse vehicular services, utilizing network slicing and hierarchical controllers to manage a dynamic resource pool while ensuring service isolation across different network segments. In~\cite{bibitem90}, a software-defined satellite-terrestrial integrated network architecture was proposed,  integrating Information-Centric Networking and SDN technologies to provide flexible management and efficient content retrieval, along with cooperative caching schemes to reduce traffic load.

Furthermore, network slicing in SD-NTNs has been intensively investigated~\cite{bibitem35,bibitem38,bibitem39,bibitem42,bibitem91,bibitem92,bibitem93,bibitem94}. In~\cite{bibitem35}, network slicing in software-defined space-air-ground integrated networks was investigated to strike the trade-off between communication and computing resource consumption. In~\cite{bibitem94}, network slicing in software-defined satellite-terrestrial integrated networks was investigated to minimize link resource utilization and the number of servers used. However, these strategies for network slicing in~\cite{bibitem35,bibitem94}do not account for the time-varying nature of SD-NTNs. In contrast, several works have considered the dynamic characteristics of SD-NTNs~\cite{bibitem38,bibitem39,bibitem42,bibitem91,bibitem92,bibitem93}. Specifically, in~\cite{bibitem38}, network slicing in software-defined space information networks was investigated, jointly utilizing communication, storage, and computing resources to maximize the number of completed missions while ensuring end-to-end latency requirements. In ~\cite{bibitem39}, network slicing in software-defined LEO satellite networks was investigated to minimize satellite-to-satellite resource consumption while efficiently serving terrestrial tasks. In~\cite{bibitem42}, network slicing in software-defined satellite-terrestrial integrated network was investigated to optimize virtual network function (VNF) deployment and routing with the goal of minimizing service latency. In~\cite{bibitem91}, network slicing in software-defined space information networks was investigated to optimize routing with service function chain constraints, aiming to maximize network flow. In~\cite{bibitem92}, group sparse network slicing in software-defined space information networks was investigated to strike the trade-off between network maximum flow and network coordination overhead. In~\cite{bibitem93}, a potential game-based network slicing in software-defined satellite edge computing was proposed to minimize deployment costs and maximize network payoff.

Focusing on radio resource management, system mobility, and onboard traffic scheduling, these studies ultimately impact NTN networking, spanning from UE access to inter-satellite connectivity. Building on this foundation and considering the latest trends in NTN development, we provide some innovative perspectives on these elements in the following sections to help advance the evolution of space Internet management.

\section{GenAI-empowered Satellite Beamforming}

\subsection{GenAI for NTNs}
Recently, Generative Artificial Intelligence (GenAI) has gained significant attention within the field of wireless communication, opening new frontiers for intelligent automation and network optimization~\cite{bibitem95}. Originally popular in applications such as natural language processing, image generation, and creative design, GenAI techniques have now started making an essential impact on wireless communication systems by enabling more efficient and adaptive resource management, interference mitigation, and network planning. These advancements come from the generative models’ ability to learn complex patterns, generate new data samples, and predict network behaviors, which are crucial for addressing the growing complexity of modern communication systems. The rapid evolution of wireless communication technologies, including the shift toward 6G and NTNs, has further fueled the integration of GenAI into the industry, allowing for more agile, data-driven decisions for radio resource allocation.

In particular, conventional DL methods have played a critical role in solving resource allocation problems by learning from historical data and predicting optimal resource usage patterns. However, they often face limitations when dealing with highly complex and changing network conditions, especially those found in NTNs, where rapid movement of satellites, fluctuating channel states, and varying traffic demands require real-time adaptability. While DL excels in classification and prediction tasks, it struggles with generating novel solutions for unseen scenarios, which are common in NTNs. GenAI, on the other hand, represents an evolution in AI's role in radio resource allocation by offering more powerful capabilities in generating new, optimized resource allocation strategies on the fly. Unlike DL, which typically relies on predefined datasets and supervised learning approaches, GenAI can autonomously create new data points, simulate network behaviors, and predict optimal resource allocations even in the absence of complete or up-to-date information. This generative capacity is critical in NTNs, where conditions change rapidly. By leveraging GenAI, NTNs can benefit from more proactive and flexible resource management, reducing latency and improving network efficiency in real-time.

Several case studies demonstrate the promise of GenAI in NTNs, showcasing its ability to manage radio resources with remarkable efficiency and adaptability. For instance, in the context of CSI estimation and beamforming design, GenAI has been leveraged to predict and generate accurate CSI in real-time, even in dynamic and unpredictable NTN environments. Traditional methods often struggle with outdated CSI due to the high mobility of satellites, but GenAI models can dynamically generate predictive CSI estimates, ensuring that resource allocation decisions are based on up-to-date information. This capability has also been applied to beamforming design, where GenAI autonomously generates optimized beamforming vectors that not only meet QoS requirements but also mitigate interference across multiple layers of the NTN. By continuously learning from real-time network data, GenAI enables faster and more efficient beamforming designs, significantly reducing the delay associated with traditional optimization methods. The following subsections will delve deeper into these use cases, highlighting the crucial role GenAI plays in advancing NTN operations.

\subsection{Channel Estimation}

\begin{figure*}[!t]
\centering
\includegraphics[width=1.4\figwidth]{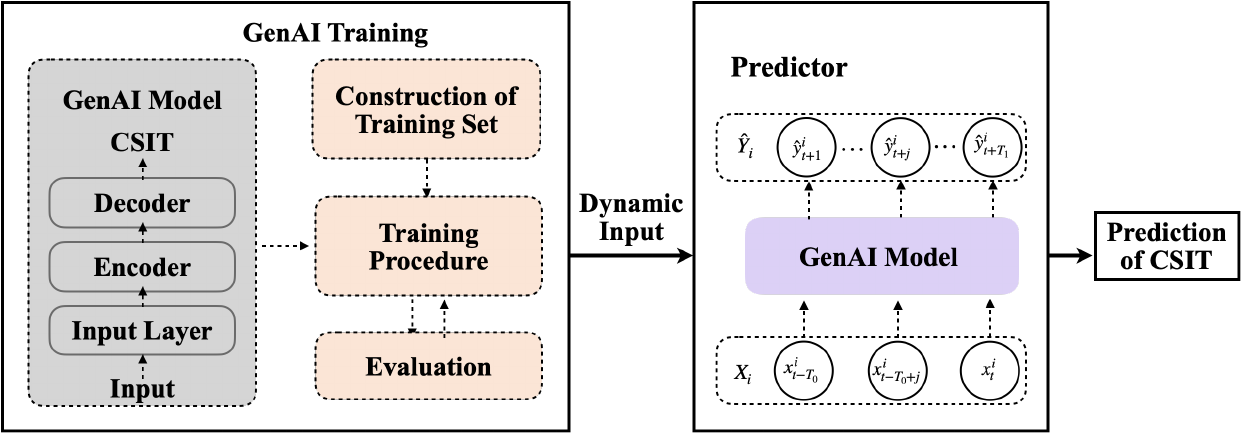}
\caption{Structure of GenAI-empowered CSI prediction.}
\label{fig2}
\end{figure*}

The imperfect CSI at the Transmitter (CSIT) originates from satellite’s mobility and report delay. To tackle this problem, the satellite could predict the long-term CSIT to avoid using outdated CSI. In particular, we predict the future CSIT according to the historical CSI samples. Given the volume of non-linear traffic data, applying a GenAI model is considered as one of the most promising approaches for CSI prediction. 

Fig. 2 illustrates a comprehensive framework for GenAI-enabled CSI estimation in NTNs. This framework leverages the data collection phase before the training process, which typically involves gathering trajectory and CSI data from previous communication scenarios. These collected data points serve as the foundation for pre-training. GenAI autonomously generates new data points, simulating a wide range of conditions to construct an enriched training dataset for various scenarios encountered in NTN communication systems. This ability to generate synthetic yet representative data allows the GenAI to address the lack of labeled datasets, particularly in highly dynamic NTNs. Once the dataset is constructed, the GenAI model undergoes a fine-tuning process. Fine-tuning helps the model adapt its parameters to the specifics of the NTN communication scenario at hand, further improving prediction accuracy.

A commonly used GenAI architecture, such as Transformer or Diffusion models, typically consists of three fundamental components: the input layer, encoder layer, and decoder layer. These components work together to facilitate the generation and prediction processes. The input layer receives the raw input data, such as past trajectory information, CSI, or even random noise, depending on the specific GenAI model used. Then, the encoder processes the input data to produce a rich and compact representation of the underlying features. In Transformers, the encoder employs mechanisms such as self-attention to capture complex relationships within the input sequence. This allows the model to understand long-range dependencies and contextual information, which is crucial for accurate prediction. In the context of Diffusion models, the encoder progressively refines noisy inputs, learning to denoise and reconstruct meaningful representations at each stage. Subsequently, the decoder layer is responsible for transforming the encoded data back into a meaningful output—in this case, CSI data. For Transformers, this involves generating sequences of data based on the encoded information and the model's internal learned representations. In Diffusion models, the decoder gradually refines the denoised representation, ultimately converging on the final predicted CSI data.

Finally, the GenAI model undergoes an evaluation phase to assess its accuracy and generalization capabilities across different NTN communication scenarios. By automating data generation and enhancing the prediction process through advanced model architectures, this framework demonstrates how GenAI can revolutionize CSI estimation, offering enhanced adaptability and performance in complex and dynamic communication environments.

\subsection{Predictive Beamforming}

\begin{figure*}[!t]
\centering
\includegraphics[width=1.5\figwidth]{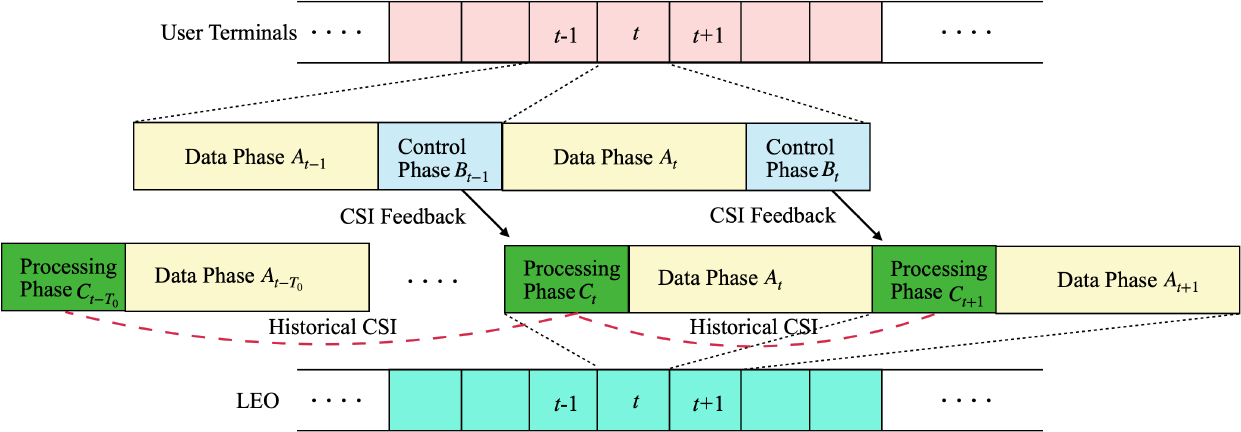}
\caption{Predictive Beamforming Protocol.}
\label{fig3}
\end{figure*}

Nevertheless, effectively processing the predictive CSI results remains a formidable challenge, introducing additional processing delays that consequently compromise the performance of NTNs. To address this problem, the predictive beamforming has been considered as a promising solution. Predictive beamforming offers a comprehensive approach by directly generating beamforming parameters from outdated CSIT, eliminating the need for separate CSI prediction and precoder design stages. This reduces signal processing overhead and improves beamforming accuracy~\cite{bibitem63}. Meanwhile, GenAI emerges as a powerful tool for tackling the complexities of this approach, as its prowess in time-series prediction and capturing non-linear relationships proves advantageous~\cite{bibitem95}.

Fig. 3 presents a comprehensive overview of proposed predictive beamforming protocol, specified for a FDD system, where channel conditions, including user deployment and channel coefficients, remain constant within each communication round. For conventional FDD protocol, each communication round t is structured into three phases: the data phase At, the control phase Bt, and the processing phase Ct. In particular, in the last communication round (t-1), the receiver estimates CSI from the received pilot signal and reports the estimated CSI to the transmitter during the control phase Bt-1. Upon receiving the reported CSI from the last round, the BS predict the CSIT in the next time slot t via the CSI prediction method and generates the beamforming design accordingly. This involves additional write/read operations better different modules, leading to additional processing delay.

To tackle this problem, we have proposed a novel predictive beamforming protocol as illustrated in Fig. 3. The primary distinction between the proposed predictive beamforming protocol and the conventional beamforming protocol lies in the processing phase, denoted as Ct. In traditional beamforming, the beamformer is predominantly generated based on the predicted CSIT in the previous communication round, Bt-1. In contrast, the predictive beamforming protocol retains the CSIT reported over the last T0 communication rounds\footnote{T0 represents the maximum number of historical CSIT entries that can be stored for each user. This constraint is crucial for limiting the storage requirements and computational complexity of the predictive beamforming protocol. The proposed protocol can generate the beamformer using any number of historical CSIT entries, provided the number does not exceed T0.}, and utilizes this accumulated historical CSIT to generate the beamformer. This enhanced context-awareness leads to better anticipation of channel variations, thereby providing a more robust solution to the problem of dynamic user mobility and fluctuating channel conditions in wireless communication networks. Moreover, it saves the write/read delay for CSI prediction module and processing module.

\section{Mobility Management Enhancement}  

In terrestrial networks, handover mechanisms enable a service BS to maintain communication when the mobile UE moves away from the current cell upon reaching cell overlap to trigger the HO process, with BSs fixed in place~\cite{bibitem65,bibitem67,bibitem96}. Conversely, in LEO-based NTN systems, multiple satellites from different orbits rapidly pass over the target area, each providing temporary cell coverage~\cite{bibitem83,bibitem97}. The serving satellite maintains communication within its coverage window until the UE is under multi-coverage for handover. Thus, traditional TN handover strategies cannot be directly applied to NTN due to its dynamic nature, prompting specific mobility management strategies for NTN.

\subsection{Seamless Handover in NTN}

\subsubsection{Distinct NTN signal variations}

\begin{figure}[!t]
\centering
\includegraphics[width=0.9\figwidth]{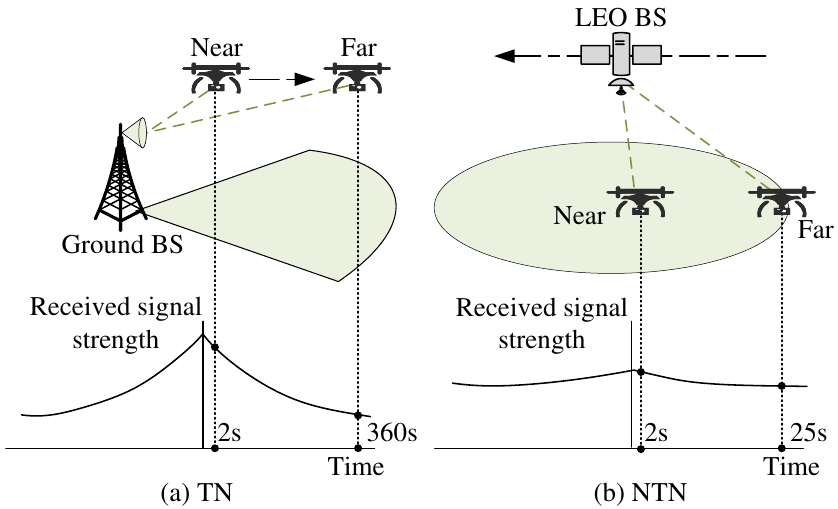}
\caption{Spatial-temporal signal variation differences between TN and NTN.}
\label{4}
\end{figure}

As depicted in Fig. 4, a UE (in this case, a UAV) moving from the center to the edge of a TN cell will experience significant signal strength reduction, primarily due to increased propagation distances~\cite{bibitem3,bibitem10}. However, signal variations in NTN are relatively minor due to satellite high altitude~\cite{bibitem98,bibitem99}. In the temporal dimension, the service time of a LEO satellite may last only several seconds to a minute. Given the extended handover signaling delays and the limited onboard capability, the time available for executing a handover in NTN is even compressed. These factors complicate the determination of the best HO target for UEs in rapidly changing environments, necessitating different link measurement and HO evaluation conditions in NTN.

\subsubsection{Pre-configurable handover in NTN}
The NTN mobility management cannot rely solely on signal quality metrics like reference signal received power (RSRP) due to the swift movement of satellites, which can affect real-time signal measurements. Yet each satellite follows a stable orbital path, allowing for predictable coverage over target areas. This predictability enables pre-calculation of multi-satellite coverage and such durations. Additionally, satellites on the same orbit pass over a target area in sequence, exhibiting similar signal quality changes. These factors offer the potential for advance handover preparations in NTN. Hence, 3GPP Rel-16 introduced the Conditional Handover (CHO) technique, which involves executing the handover preparation phase preemptively and then monitoring the conditions of all candidate BSs~\cite{bibitem10}. Once a candidate meets the HO conditions, only a final Detach \& Synchronize step with the source BS is needed~\cite{bibitem10}. CHO is designed to reduce HO failures in dynamic scenarios, making it possible for NTN mobility enhancements~\cite{bibitem96,bibitem100}. It is natural for UE in NTN to use satellite motion information (e.g., satellite ID, ephemeris data) as the measurement condition for a location-based CHO strategy. However, relying solely on this condition could cause UEs to trigger handovers to low-quality cells. For fast-moving satellites, the time of stay is also a crucial metric to ensure NTN service stability and minimize the HO number. Similarly, since satellites passing over a target area create a signal quality pattern that rises and then falls, the change in signal quality is also a vital HO evaluation condition. Hence, the monitoring condition should combine several types of metrics, but all should account for predictable satellite movement to improve network performance.

\subsubsection{NTN handover configuration}
Orbital satellite movements facilitate not just CHO but also the prediction of potential future handover targets, establishing a pre-configured NTN handover sequence, as studied in~\cite{bibitem87,bibitem101}. For a LEO constellation, the coverage status over a target area within a specific time duration is predictable. Thus, the satellites that UES in the area connect to in this period is an ordered sequence temporally. Due to the dynamics, managing a high HO rate is essential in NTN, often leading to frequent HO signaling storms. Pre-configured handover sequence can reduce signaling interactions between satellites and UEs and between satellites themselves, enhancing the HO success rate. Moreover, a pre-configured NTN handover sequence can ensure better service continuity. Traditional HO strategies sequentially select the best target when under multiple coverage but may overlook the dependence between successive handovers, impacting network service continuity. This issue can be mitigated by configuring the NTN handover sequence, which can be integrated with resource allocation, time-varying network awareness, and network slicing to enhance NTN service capabilities.

\subsection{Mobility Support Between TN And NTN}

\begin{figure}[!t]
\centering
\includegraphics[width=\figwidth]{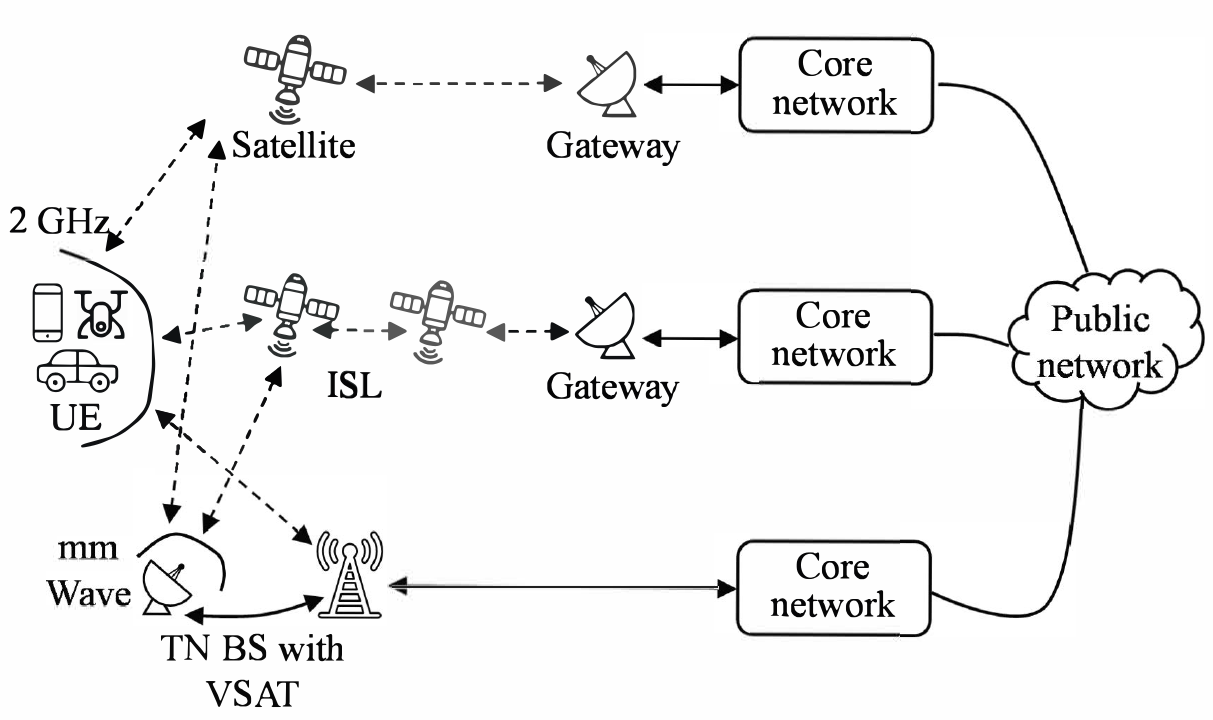}
\caption{UE connectivity and BS signaling in integrated TN-NTN.}
\label{5}
\end{figure}

\subsubsection{Integrated TN-NTN architecture}
Since the inception of 5G and beyond, the integration of TN and NTN has played a crucial role in the evolution of mobile systems~\cite{bibitem5,bibitem8,bibitem88}. As shown in Fig. 5, satellites now can act as space stations managed under a unified spectrum with ground BSs, capable of signal interaction and jointly managed backhaul by network operators~\cite{bibitem102}. In the fronthaul, UEs perceive co-existed TN and NTN cells and can select either segment for access based on signal conditions and their own requirements, which necessitates unified TN-NTN mobility management~\cite{bibitem12,bibitem103}. In the integrated system, NTN's role extends beyond merely providing network connectivity in areas beyond the reach of TN. It is increasingly valued for complementing network capacity, especially in sparse TN distributions or when UE moving to cell edges, thus improving overall service capabilities~\cite{bibitem2}. This integration is particularly beneficial for mobile UEs across regions. For instance, in the increasingly emphasized low-altitude economy, UAVs traveling from urban to rural areas need to maintain a constant connection with the mobile network for real-time reception of task-specific payloads and control commands. Current TN may only ensure network coverage in urban. An integrated TN-NTN system ensures complete connectivity coverage along the entire UAV corridor, potentially operating in a service model that transitions from TN to a mix of TN-NTN, and then solely to NTN. With this integration, it is introduced new challenges in efficient mobility management between TN and NTN to ensure the reliability and continuity of data services.

\subsubsection{Unified assessment for TN-NTN}
UEs in integrated TN-NTN experience uninterrupted data service through seamless handovers between TN and NTN segments~\cite{bibitem3,bibitem99}. However, as illustrated in Fig. 4, distinct variations in signal strength between TN and NTN, both temporally and spatially. Spatially, TN signal variations primarily depend on the propagation distance from cell center to edge. In contrast, NTN signals are mainly affected by the elevation angle that affects the line-of-sight (LoS) probability and shadow fading, with lower angles inhibiting valid satellite communications. Temporally, the rapid movement of satellites may necessitate multiple NTN handovers within the duration of a single TN cell's service. To address these differences, the pre-configured NTN HO sequence mentioned above is beneficial, aligning with the TN service duration to uniformly assess HO candidates between TN and NTN. It is expected for each UE to receive high service rates while reducing HO overhead for mobility support. Hence, the HO condition should jointly consider the SINR, time of stay, and signaling dynamics for both TN and NTN candidates. Additionally, it should consider the service capability comparison between TN and NTN, with NTN not only supplementing but also potentially competing with TN to ensure a sustainable and capable data service upon switching segments.

\subsubsection{Handover between TN-NTN}
When mobile UE under heterogeneous HO candidate coverage, the TN-NTN handover mechanism activates~\cite{bibitem99}. The serving BS sends handover request to both TN and NTN candidates. For satellite candidate, it computes the NTN handover sequence for UE based on predictable satellite movements for comparison with the TN BS candidate. UE then assesses the handover benefits of TN and NTN under unified conditions to select the most suitable next service segment. It is important to note that pre-configuring the NTN serving sequence involves reserving resources in subsequent satellites, a process that incorporates constellation management techniques. This is crucial when managing large-scale LEO constellations to calculate the best NTN serving sequence for UEs with diverse requirements while maintaining efficient network resource allocation. Therefore, mobility management in integrated TN-NTN systems must consider not only dynamic access between UEs and BSs but also synchronize with resource management across the heterogeneous network to achieve joint TN-NTN optimization. This technique aligns with the 6G vision and future network developments aiming for TN-NTN unification.
\section{Network Slicing for SD-NTN}

\subsection{SD-NTN Architecture}

\begin{figure*}[!t]
\centering
\includegraphics[width=1.4\figwidth]{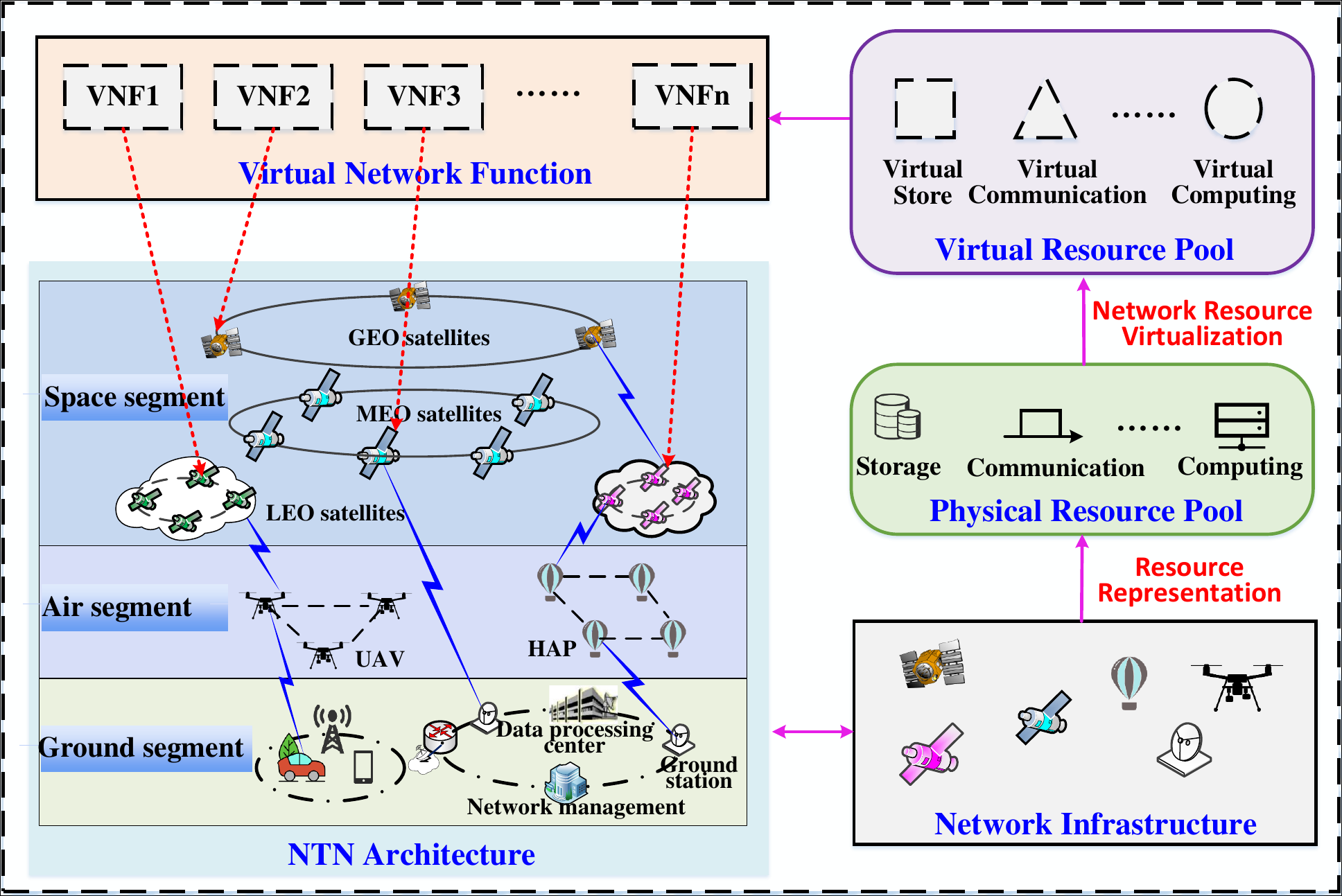}
\caption{SD-NTN architecture.}
\label{6}
\end{figure*}

The SD-NTN architecture employs virtualization technologies to convert physical network resources—such as communication, storage, and computing—across space, air, and ground layers into virtual resources~\cite{bibitem89}, as depicted in Fig. 6. These resources are pooled into a unified resource pool that supports networking across diverse network segments~\cite{bibitem37,bibitem89}. The unification facilitates dynamic resource allocation and optimization, tailored to meet the varied demands of different services. Key to this architecture are two pivotal virtualization technologies, SDN and NFV, which facilitate effective resource management and agile network orchestration~\cite{bibitem35,bibitem36,bibitem89}: 
\begin{itemize}
    \item[1.]	\emph{Software-Defined Networking (SDN):} SDN decouples the control plane from the data plane, enabling centralized and programmable control over the entire network~\cite{bibitem41,bibitem104}. In the SD-NTN architecture, SDN provides a centralized view of both physical and virtual resources across space, air, and ground segments, allowing network operators to dynamically manage and optimize resource allocation in real-time~\cite{bibitem34,bibitem37,bibitem52}. This centralized control is crucial for managing the highly dynamic and distributed nature of non-terrestrial networks, where satellite constellations, HAPs, UAVs, and terrestrial nodes need to be coordinated seamlessly. Additionally, SDN controllers can monitor network conditions, such as link availability, bandwidth usage, and node health~\cite{bibitem34,bibitem89}. By enabling programmable control, SDN enhances network flexibility, allowing it to dynamically adjust flow routing and efficiently adapt to varying traffic demands and frequent topology changes~\cite{bibitem34,bibitem43}. It also simplifies QoS management, strengthens network security, and supports rapid fault recovery, further enhancing overall network scalability and flexibility~\cite{bibitem34,bibitem43,bibitem52}. 
    \item[2.]\emph{Network Function Virtualization (NFV):} NFV decouples network functions from dedicated physical hardware, allowing them to be flexibly deployed as software instances, referred to as virtual Network Functions (VNFs), on various physical nodes in the NTN~\cite{bibitem40,bibitem105}. In the SD-NTN architecture, this decoupling enables the flexible deployment of VNFs on satellites, HAPs, UAVs, or ground stations, depending on the service demands and resource availability~\cite{bibitem46}. This flexibility allows network operators to scale and optimize network functions without requiring specialized hardware upgrades, thereby reducing operational costs and simplifying network maintenance~\cite{bibitem106,bibitem107}. Furthermore, NFV enhances efficient resource utilization by allowing multiple VNFs to share the same physical node, improving network adaptability to evolving requirements of diverse services~\cite{bibitem38}. NFV also increases network agility by facilitating the rapid deployment and migration of VNFs across different nodes in the NTN~\cite{bibitem107}. This dynamic allocation and redeployment of VNFs ensure that SD-NTNs can maintain high levels of service availability and reliability, even under changing network conditions~\cite{bibitem107}. 
\end{itemize}
By integrating NFV and SDN technologies, SD-NTNs achieve flexible, scalable, and efficient resource management across the network~\cite{bibitem34,bibitem37,bibitem43,bibitem44}, ultimately improving the adaptability and scalability of NTNs to handle diverse and complex tasks.

\subsection{Network Slicing}

\begin{figure}[!t]
\centering
\includegraphics[width=\figwidth]{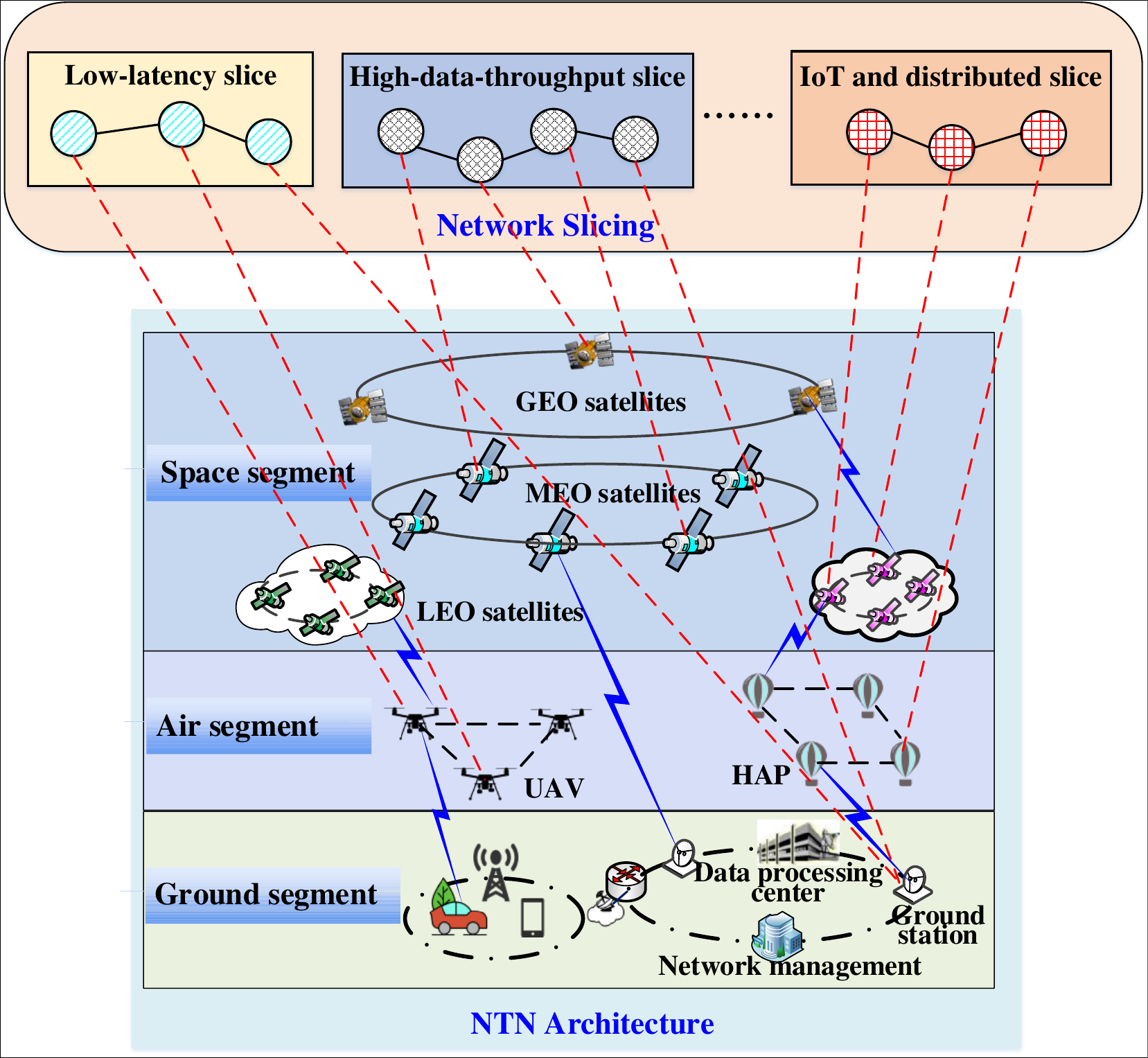}
\caption{Illustration of network slicing in SD-NTN.}
\label{7}
\end{figure}

In SD-NTNs, network slicing plays a pivotal role in enabling flexible, scalable, and efficient service provisioning, allowing the network to meet diverse service requirements over a shared physical infrastructure~\cite{bibitem44,bibitem45,bibitem46}. Specifically, network slicing refers to the creation of multiple isolated virtual networks, or slices, that can operate simultaneously over the same physical infrastructure~\cite{bibitem44,bibitem45,bibitem46}, as shown in Fig. 7. Each network slice is customized to meet the specific requirements of different services. By creating distinct virtual networks, SD-NTNs ensure that each slice has dedicated resources and configurations optimized for specific service requirements.

SD-NTNs enables VNFs to be flexibly deployed on different physical nodes, enabling each slice to meet the specific requirements of each service~\cite{bibitem35,bibitem38,bibitem46}. Each requested service can be represented by a Service Function Chain (SFC), which consists of a sequence of VNFs in the predefined order~\cite{bibitem35,bibitem38,bibitem108,bibitem109}. To complete the requested service, the mission flow must traverse each VNF in the specified sequence defined by the SFC~\cite{bibitem35,bibitem38,bibitem108,bibitem109}. Therefore, in SD-NTNs, there are two key technologies essential for network slicing: the first is VNF deployment, which involves strategically placing virtual functions on suitable physical nodes to maximize service performance~\cite{bibitem35}; the second is flow routing, which requires networking schemes in SD-NTNs that meet both the SFC constraints and the multi-dimensional resource constraints~\cite{bibitem35,bibitem38}. 

\subsubsection{Efficient VNF deployment}
Different VNF deployment strategies are crucial for optimizing network performance in SD-NTNs~\cite{bibitem35,bibitem38,bibitem92}, which must align with the specific requirements of each service type, such as low-latency, high-data-throughput, and IoT or distributed services. The following are several VNF deployment cases based on the service characteristics.
\begin{itemize}
    \item[1.] \emph{Low-latency services:} For network slices with strict low-latency requirements, such as those supporting URLLC services, VNFs should be deployed as close to the end-users as possible. In SD-NTNs, this can be achieved by deploying VNFs on HAPs, UAVs, or ground nodes near the end-users, significantly reducing transmission distance and minimizing propagation delays. This VNF deployment strategy ensures that latency-sensitive services, such as autonomous driving and remote medical procedures, maintain the ultra-low latency necessary for real-time data processing and decision-making.
    \item[2.] \emph{High-data-throughput services:} For network slices supporting services that require high data throughput, such as eMBB services, VNFs should be deployed on physical nodes with significant processing, storage, and bandwidth capabilities. In SD-NTNs, this can be achieved by deploying VNFs on nodes with higher communication capacity, such as MEO or GEO satellites, where larger amounts of data can be transmitted more efficiently across vast areas. Additionally, deploying VNFs on central data centers or ground stations with abundant computing resources allows the network to handle data-intensive services, ensuring large-scale data transmissions occur without congestion or bottlenecks. This strategy supports high-throughput applications like video streaming, augmented reality (AR), and cloud-based services, ensuring minimal congestion and maintaining high-performance levels.
    \item[3.] \emph{IoT and distributed services:} For network slices supporting IoT and distributed services, such as mMTC services, VNFs should be deployed on nodes optimized for scalability and energy efficiency. In SD-NTNs, this can be achieved by deploying VNFs on LEO satellites, UAVs, or edge nodes that can support a large number of low-data-rate IoT devices. These nodes distribute processing closer to the devices, reducing latency and alleviating congestion in the core networks. By deploying VNFs across multiple nodes, the network can efficiently manage the high-density connectivity required by IoT devices, while optimizing resource allocation and minimizing energy consumption. This strategy supports seamless operation for applications such as smart city infrastructure, industrial automation, and environmental monitoring in vast, decentralized networks.
\end{itemize}

\subsubsection{SD-NTN flow routing}
In SD-NTNs, the mission flow must traverse each VNF in the specified sequence defined by the SFC~\cite{bibitem35,bibitem38,bibitem108,bibitem109}. However, due to the high-speed movement of satellites, UAVs, and HAPs, the connectivity between nodes in SD-NTNs is intermittent, resulting in time-varying topology~\cite{bibitem44,bibitem45,bibitem49,bibitem50,bibitem51}. Therefore, flow routing strategies satisfying SFC constraints, designed for static network topologies, are not directly applicable to time-varying SD-NTNs~\cite{bibitem38,bibitem42,bibitem91}. In addition, due to the size, weight, and cost limitations of satellites, UAV, and HAPs, the payload they can carry are strictly limited~\cite{bibitem45,bibitem51,bibitem52}. Furthermore, in SD-NTNs, fulfilling complex service requests typically requires the collaboration of multi-dimensional heterogeneous resources, including communication, storage, and computing, across different nodes. Therefore, in time-varying SD-NTNs, it is crucial to design optimized flow routing strategies that not only meet SFC constraints but also ensure the efficient allocation of limited communication, storage, and computing resources. These strategies must consider the dynamic and intermittent connectivity between nodes to maximize resource utilization and improve network performance. One approach is to utilize the multi-functional time expanded graph (MF-TEG) to model the time-varying SD-NTN topology with communication, storage, and computing resources~\cite{bibitem110}. MF-TEG can be further utilized to design VNF deployment and flow routing strategies to ensure efficient utilization of network resources and improve overall network performance.

\section{Future Research and Open Problem}

\subsection{Future Research}

\subsubsection{Interference management in NTNs}
While the aforementioned studies have addressed intra-system interference within NTNs, the broader challenges of managing inter-system interference remain unresolved. In 6G system, the co-existence of LEO satellites and UAVs in the same frequency bands transforms inter-system interference into a significant challenge. Some early work relied on cognitive radio approaches for inter-system interference migration, typically designating one system as primary and the other as secondary~\cite{bibitem111,bibitem112,bibitem113}. These approaches aimed to maximize the throughput of the primary system while guaranteeing the QoS for the secondary. However, these efforts suffer from inherent limitations. By simply adopting intra-system interference management strategies like NOMA and 1-layer RSMA for inter-system scenarios, they fail to fully exploit the potential of NTNs. Consequently, they often achieve suboptimal performance, either by prioritizing the primary system while neglecting the secondary's potential or by constraining the secondary's capabilities to minimize interference. This binary approach ultimately hinders the true potential of both systems and falls short of addressing the intricacies of inter-system interference in NTNs. To tackle these problem, an effective and efficiency interference management system would be design in future research. 

\subsubsection{Fine-tuned TN-NTN unification}
The unification of TN and NTN goes beyond integrated architectures and connectivity management, extending to fine-grained orchestration of applications and services. Due to the ubiquitous coverage of NTN, UEs typically experience co-existed coverage from both TN and NTN. In such scenarios, NTN must play the role of service availability and reliability reinforcement, which is particularly valuable in applications such as highway vehicular networks. Additionally, as BSs support multicast-broadcast services (MBS), NTN can enable service scalability by leveraging its natural wide coverage to provide efficient multicast/broadcast for data delivery cooperated with TN~\cite{bibitem114,bibitem115}. Thus, beyond merely supplementing connectivity, NTN and TN enhance overall system performance through strategic orchestration, leveraging their respective strengths.

\subsubsection{Reliability-enhanced network slicing}
In SD-NTNs, ensuring network reliability presents significant challenges due to the network's heterogeneous composition, which includes satellites, UAVs, HAPs, and ground stations. These nodes are prone to disruptions caused by various factors such as hardware failures, link interruptions, environmental conditions, or mobility-induced disconnections. For example, satellites may face malfunctions or orbital adjustments, while UAVs and HAPs could experience power loss or weather-related disturbances. Ensuring that network slices remain resilient and can quickly recover from such failures is crucial for maintaining service continuity in SD-NTN. Future research should focus on developing robust, fault-tolerant mechanisms that can detect and respond to failures in real time, ensuring minimal service disruption. These strategies include dynamic reconfiguration of VNFs, allowing services to migrate seamlessly across available nodes, such as satellites, UAVs, and HAPs. In addition, real-time fault detection strategies are essential for improving the overall reliability of SD-NTNs. These strategies should continuously monitor network conditions and predict potential failures before they occur. By integrating predictive models and machine learning techniques, the network can proactively identify early signs of hardware malfunctions, link instability, or resource exhaustion. This enables preemptive actions, such as reconfiguring network slices or adjusting resource allocation, to prevent failure and enhance the overall reliability of the network.

\subsection{Open Problems}

\subsubsection{Integrated sensing and communications in NTNs}
The integration of Sensing and Communication (ISAC) in NTNs represents a highly promising advancement that holds the potential to revolutionize both communication and sensing capabilities in spaceborne and airborne platforms. By optimizing spectrum, power, and hardware for dual functionality, ISAC enables NTNs to support tasks like environmental monitoring, object detection, and surveillance within a unified framework~\cite{bibitem95,bibitem116}. Additionally, ISAC can drive NTN advancements in autonomous navigation, disaster management, and real-time situational awareness, significantly improving operational efficiency. However, ISAC faces major challenges in NTNs due to the dynamic propagation environment. Frequent handovers and long distances increase latency and complicate synchronization, while limited power further constrains ISAC operations. Achieving effective ISAC in NTN will require advanced signal processing and adaptive beamforming techniques to manage these complexities.

\subsubsection{TN-NTN competitive integration}
The deployment and service of TN and NTN differ significantly in spatial-temporal aspects, including distribution patterns, fixed vs. dynamic BSs, coverage status, and cell service durations~\cite{bibitem8,bibitem85}. When arranging both TN and NTN, operators must swiftly design transmission schemes and manage mobility for the heterogeneous system. A key challenge is the strategic assignment of transmission tasks for TN and NTN. Theoretically, any task that can be transmitted via TN could also be offloaded to NTN, given the global coverage. When TN and NTN alternately have transmission advantages, there must be a foresighted strategy for task division and joint transmission. This may benefit from advanced computational and analytical technologies, such as federated learning and digital twins, to ensure efficient data delivery.

\subsubsection{Security and privacy in NTN slicing}
In SD-NTNs, enhancing security and privacy across network slices is a significant challenge due to the diverse cyber threats inherent in the distributed architecture. These networks are particularly vulnerable to security risks such as eavesdropping, data tampering, and denial-of-service attacks, exacerbated by shared infrastructure which increases the risk of cross-slice attacks and data exposure. Moreover, the time-varying topology of SD-NTNs further complicates real-time security measures. It’s necessary to develop secure isolation mechanisms that ensure data and control flows remain segregated even on shared resources, along with advanced encryption methods, authentication protocols, and secure resource management strategies for the network dynamics. Furthermore, slice-specific security policies should be dynamically applied, ensuring that security measures adapt to the unique requirements of each slice. For example, slices handling URLLC services may require more stringent security protocols compared to slices designed for low-latency entertainment services.
\section{Conclusions}

In this paper, we explored the critical role of non-terrestrial network (NTN) deployment in the evolution of 6G mobile systems. Given the extensive coverage and rapid dynamics of LEO satellites, we reviewed the unique challenges they pose to NTN networking, including the development and issues related to radio access, satellite handover, and network slicing mechanisms. Furthermore, we introduce some innovative perspectives on managing NTN radio resources, system mobility, and onboard traffic scheduling, aiming to achieve efficient, high-quality, and reliable global NTN networking and data services. Finally, we identified several open problems and highlighted useful techniques and directions for advancing the deployment of space Internet. It is hoped that this study will provide insights that will be valuable for the design, operation, and optimization of NTNs.



\end{document}